\documentclass[english,superscriptaddress,floatfix]{revtex4}
\usepackage[T1]{fontenc}
\usepackage[latin9]{inputenc}
\usepackage[pdftex]{color}
\usepackage{babel}
\usepackage{amsmath}
\usepackage{amssymb}
\usepackage{esint}
\usepackage[unicode=true,
 bookmarks=false,
 breaklinks=true,pdfborder={0 0 1},backref=section,colorlinks=true]
 {hyperref}
\usepackage{graphicx}
\hypersetup{
 pdfcreator={},pdfproducer={LaTeX with hyperref},linkcolor=blue,anchorcolor=blue,citecolor=blue,filecolor=red,menucolor=red,pagecolor=red,urlcolor=blue,pdfstartview=FitV,pdfhighlight=/I,pdfpagelayout=OneColumn,hypertexnames=true}

\makeatletter

\@ifundefined{textcolor}{}
{%
 \definecolor{BLACK}{gray}{0}
 \definecolor{WHITE}{gray}{1}
 \definecolor{RED}{rgb}{1,0,0}
 \definecolor{GREEN}{rgb}{0,1,0}
 \definecolor{BLUE}{rgb}{0,0,1}
 \definecolor{CYAN}{cmyk}{1,0,0,0}
 \definecolor{MAGENTA}{cmyk}{0,1,0,0}
 \definecolor{YELLOW}{cmyk}{0,0,1,0}
}

\usepackage{babel}

 \@ifundefined{textcolor}{}{%
  \definecolor{BLACK}{gray}{0}
  \definecolor{WHITE}{gray}{1}
  \definecolor{RED}{rgb}{1,0,0}
  \definecolor{GREEN}{rgb}{0,1,0}
  \definecolor{BLUE}{rgb}{0,0,1}
  \definecolor{CYAN}{cmyk}{1,0,0,0}
  \definecolor{MAGENTA}{cmyk}{0,1,0,0}
  \definecolor{YELLOW}{cmyk}{0,0,1,0}
 }

 \@ifundefined{textcolor}{}{
  \definecolor{BLACK}{gray}{0}
  \definecolor{WHITE}{gray}{1}
  \definecolor{RED}{rgb}{1,0,0}
  \definecolor{GREEN}{rgb}{0,1,0}
  \definecolor{BLUE}{rgb}{0,0,1}
  \definecolor{CYAN}{cmyk}{1,0,0,0}
  \definecolor{MAGENTA}{cmyk}{0,1,0,0}
  \definecolor{YELLOW}{cmyk}{0,0,1,0}
 }
 \@ifundefined{definecolor}{
 }{}
 \@ifundefined{definecolor}{color}{}

\usepackage{color}

\newcommand{\be}{\begin{equation}}
\newcommand{\ee}{\end{equation}}
\newcommand{\bea}{\begin{eqnarray}}
\newcommand{\eea}{\end{eqnarray}}
\newcommand{\bse}{\begin{subequations}}
\newcommand{\ese}{\end{subequations}}

\newcommand{\add}[1]{\textcolor{black}{#1}}
\usepackage{color}% define colors
\definecolor{d_red}{cmyk}{0.00, 0.81, 1.00, 0.27}
\definecolor{d_orange}{cmyk}{0.00, 0.33, 1.00, 0.00}
\definecolor{d_blue}{cmyk}{0.78, 0.47, 0.00, 0.20}
\definecolor{d_lgreen}{cmyk}{0.07, 0.00, 0.79, 0.29}
\definecolor{d_green}{cmyk}{0.66, 0.00, 0.71, 0.56}
\definecolor{d_blue}{cmyk}{0.78, 0.47, 0.00, 0.20}
\definecolor{d_dblue}{cmyk}{0.91, 0.79, 0.00, 0.22}
\definecolor{d_pink}{cmyk}{0.0, 0.79, 0.37, 0.29}
\definecolor{d_purple}{cmyk}{0.16, 0.54, 0.00, 0.70}
\definecolor{d_paleblue}{cmyk}{0.669, 0.338, 0.00, 0.373}
\definecolor{d_dpaleblue}{cmyk}{0.441, 0.290, 0.00, 0.580}
\definecolor{d_brown}{cmyk}{0.0, 0.490, 0.930, 0.350}
\definecolor{d_turquoise}{cmyk}{0.630, 0.04, 0.0, 0.440}
\definecolor{KIT-green}{RGB}{0, 150,130}
\definecolor{KIT-blue}{RGB}{70,100,170}

\def\bmx{\begin{pmatrix}}
\def\emx{\end{pmatrix}}

\usepackage[figure,table]{hypcap}% to correct a problem with hyperref

\makeatother

\begin{document}
\title{Integer and fractionalized vortex lattices and off-diagonal long-range
order}
\author{Michael A. Rampp }
\affiliation{Institut f\"{u}r Theorie der Kondensierten Materie, Karlsruher Institut
f\"{u}r Technologie, 76131 Karlsruhe, Germany}
\author{J\"{o}rg Schmalian}
\affiliation{Institut f\"{u}r Theorie der Kondensierten Materie, Karlsruher Institut
f\"{u}r Technologie, 76131 Karlsruhe, Germany}
\affiliation{Institut f\"{u}r QuantenMaterialien und Technologien, Karlsruher Institut
f\"{u}r Technologie, 76344 Karlsruhe, Germany}
\begin{abstract}
We analyze the implication of off-diagonal long-range order (ODLRO)
for inhomogeneous periodic field configurations and multi-component
order parameters. For single component order parameters we show that
the only static, periodic field configuration consistent with ODLRO
is a vortex lattice with integer flux in units of the flux quantum
in each unit cell. For a superconductor with $g$ degenerate components, fractional
vortices are allowed. Depending on the precise order-parameter manifold, they tend to occur in units of $1/g$ of the flux quantum.
These results are well known to emerge from the Ginzburg-Landau or BCS
theories of superconductivity. Our results imply that they are valid
even if these theories no-longer apply. Integer and fractional vortex lattices are transparently seen to emerge as a consequence of the macroscopic coherence and single valuedness of the condensate.
\end{abstract}
\maketitle

\section{Introduction}

The Meissner effect~\cite{Meissner1933} and the quantization of the
magnetic flux in multiply connected samples~\cite{Deaver61,Doll61}
belong to the most fundamental aspects of superconductivity. These
phenomena follow from the phenomenological Ginzburg-Landau theory~\cite{Ginzburg1950}
and from the microscopy theory of superconductivity developed by Bardeen,
Cooper, and Schrieffer (BCS)~\cite{Bardeen57_1,Bardeen57_2}. They
are, however, phenomena that occur beyond the regime of applicability
of the BCS theory. This was anticipated by London, based on the concept
of macroscopic coherence~\cite{London50}. The formal framework to
demonstrate that these phenomena are caused by a coherent condensate
and are valid more generally was provided by C. N. Yang~\cite{Yang62}
who analyzed the two-particle density matrix

\begin{equation}
\rho_{\alpha\beta\gamma\delta}^{\left(2\right)}\left(\boldsymbol{r}_{1},\boldsymbol{r}_{2};\boldsymbol{r}_{3},\boldsymbol{r}_{4}\right)=\left\langle \psi_{\alpha}^{\dagger}\left(\boldsymbol{r}_{1}\right)\psi_{\beta}^{\dagger}\left(\boldsymbol{r}_{2}\right)\psi_{\gamma}\left(\boldsymbol{r}_{3}\right)\psi_{\delta}\left(\boldsymbol{r}_{4}\right)\right\rangle \label{eq:rho2}
\end{equation}
of a many-fermion system. Here $\psi_{\alpha}^{\dagger}\left(\boldsymbol{r}\right)$
and $\psi_{\alpha}\left(\boldsymbol{r}\right)$ are fermion creation
and annihilation operators at position $\boldsymbol{r}$ and with
spin $\alpha$, respectively. Yang generalized the concept of off-diagonal
long-range order (ODLRO), initially proposed for interacting bosons
by Penrose and Onsager~\cite{Penrose51,Penrose56}, to fermionic systems
and demonstrated that ODLRO implies flux quantization with elementary
flux $\Phi_{0}=\tfrac{hc}{2e}$. The beauty of the result is that
it can be made without reference to the Hamiltonian and merely relies
of the presence of a macroscopic pair condensate. More recently, it
was shown in Refs.~\cite{Sewell90,Nieh95} that a homogeneous magnetic
field cannot exist in the bulk of a charged system, i.e. $\boldsymbol{B}=\boldsymbol{0}$
if it is spatially constant and if we ignore surface effects. This
amounts to the Meissner effect as it occurs in type-I superconductors.
In Ref.~\cite{Au1995} the argumentation was then generalized to inhomogeneous
fields with cylindrical symmetry and fields that are slowly varying
in space. 

In this paper we generalize previous conclusions that follow from
ODLRO with regards to two aspects. On the one hand, we consider periodic
magnetic fields without the restriction of slow variation in space.
We show that the only static, periodic field configuration consistent
with superconductivity is a vortex lattice with integer flux in each
unit cell. On the other hand, we consider multi-component superconducting
states and find that for a superconductor with $g$-component
order parameter the elementary flux quantum changes to $\Phi_{0}\rightarrow\frac{1}{g}\tfrac{hc}{2e}$. Hence,
 fractional vortices and fractional vortex lattices become possible. Both
results are known within the regime of validity of the Ginzburg-Landau
and BCS approaches. The former corresponds, of course, to Abrikosov's
vortex lattice of the mixed state~\cite{Abrikosov1957,Kleiner1964,Eilenberger1964,Brandt1997},
while fractional vortices were discussed in the context of superfluid
$^{3}$He~\cite{Salomaa1985}, two-gap superconductors~\cite{Babaev2002,Babaev2004,Babaev2009},
$p_{x}\pm ip_{y}$ triplet superconductors~\cite{Kee2001,Babaev2005,DasSarma2006,Chung2007,Chung2009,Vakaryuk2011,Ramachandhran2012},
or spin-orbit-coupled Bose-Einstein condensates~\cite{Ramachandhran2012}.
The composite of a half-flux vortex and the Majorana fermions bound
at its core led to significant interest given the resulting non-Abelian
fractional statistics~\cite{Read2000,Ivanov2001,Stern2004,Stone2006}.
Experimentally, Abrikosov vortex lattices were observed via small-angle
neutron diffraction~\cite{Cribier1964} and the Bitter decoration technique~\cite{Essmann1967}
in the 1960s. Evidence for fractional vortices is much sparser. In
superfluid $^{3}$He in a porous medium vortices with half the quantum
unit of fluid flow have indeed been generated in the laboratory~\cite{Autti2016} \add{and single fractional vortices have been observed in two-gap superconductors~\cite{Tanaka2018,Tanaka2021,Pina2014} in which a lattice might be stabilized by a periodic pinning array~\cite{Lin2013}}.
The extreme vortex pinning in the non-centrosymmetric superconductor
CePt$_{3}$Si was also interpreted in terms of fractionalized vortices~\cite{Miclea2009},
but unambiguous evidence for a fractionalized vortex lattice does
not exist thus far, even though there are strong arguments to expect
such a state in triplet superconductors at high magnetic field~\cite{Chung2009}. Moreover, fractional vortices may also form a vortex lattice with a non-trivial unit cell consisting of multiple fractional defects that add up to an integer flux~\cite{Volovik2022}.

Our ODLRO analysis shows that these established results  do not
rely on the validity of the Ginzburg-Landau and BCS theories. They reveal, using rather straightforward reasoning, that integer and fractional vortex lattices are tied to macroscopic coherence and the single valuedness of the condensate. On the other hand,
we can only make statements about what is quantum mechanically allowed,
not what is energetically most stable.

\section{Summary of off-diagonal long-range order}

\begin{figure}
    \centering
    \includegraphics[width = .95\textwidth]{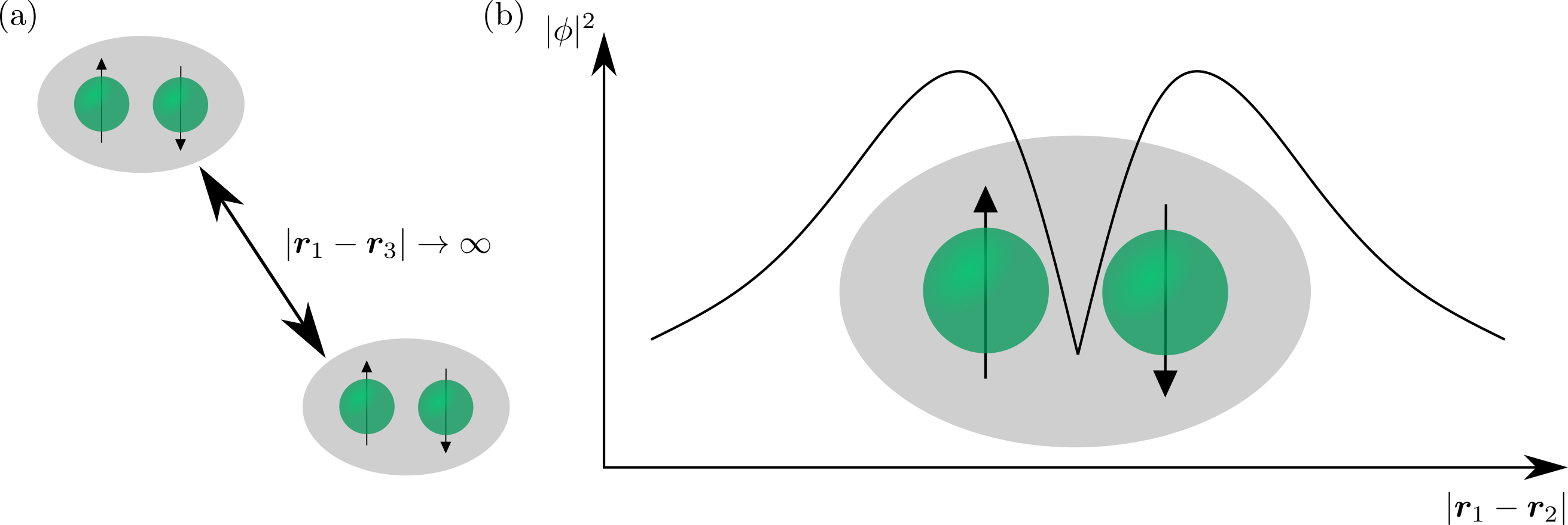}
   \caption{ODLRO in the two-particle density matrix is defined as the presence of one (or multiple) eigenvalues that scale linearly with system size. (a) As a result, the amplitude of creating and destroying two pairs of electrons is non-zero even in the limit of infinite separation of the pairs. (b) This can be understood as the presence of a macroscopic two-particle wave function describing the superconducting condensate.}
    \label{fig:2pwf}
\end{figure}

\begin{figure}
    \centering
    \includegraphics[width = .95\textwidth]{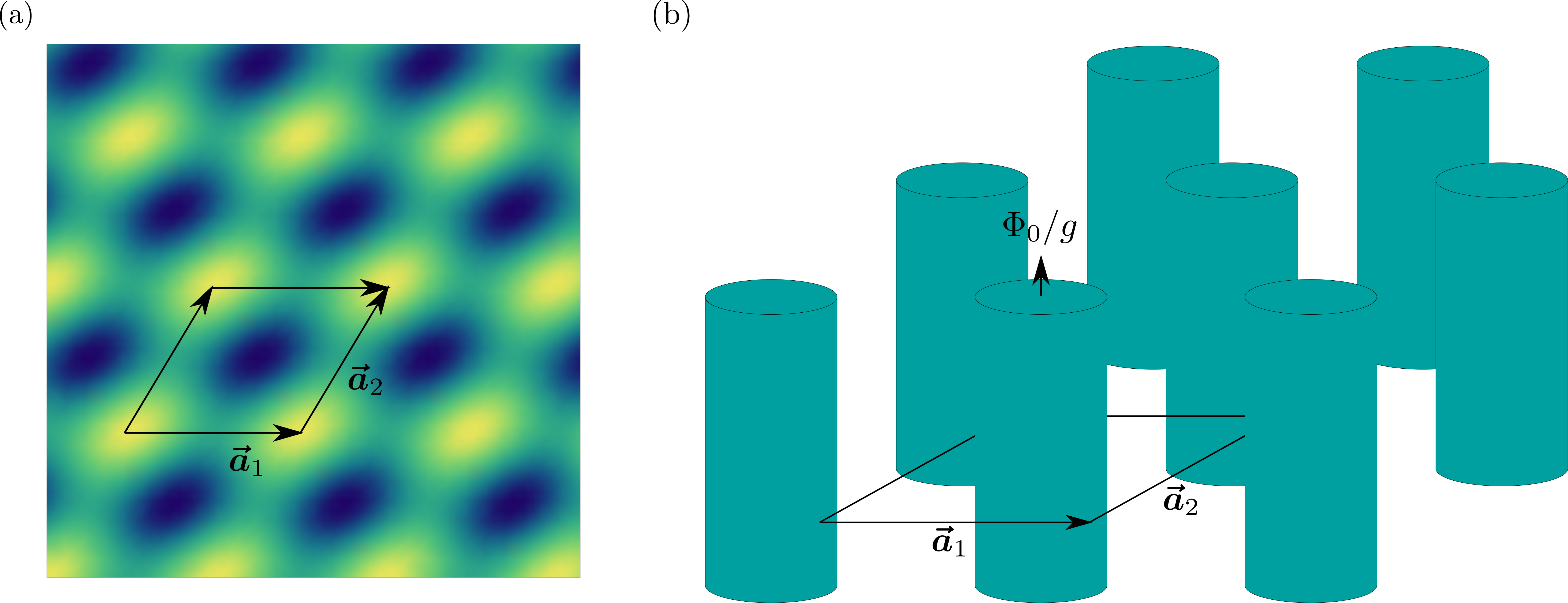}
   \caption{Renditions of a flux lattice with primitive lattice vectors $\bm{a}_1$ and $\bm{a}_2$. The quantization condition follows from requiring that the wavefunction does not depend on the order of translations, and from a less stringent requirement in the multi-component case.}
    \label{fig:Summary}
\end{figure}

We first summarize some of the main aspects of ODLRO for fermionic
systems. This brief summary follows closely Refs.~\cite{Yang62,Sewell90,Nieh95}.
We analyze the density matrix $\rho^{\left(2\right)}$ of Eq.~\ref{eq:rho2}
and consider the combined two-particle coordinates $\left(\boldsymbol{r}_{1},\alpha,\boldsymbol{r}_{2},\beta\right)$
and $\left(\boldsymbol{r}_{3},\gamma,\boldsymbol{r}_{4},\delta\right)$.
The matrix structure of interest should be understood  with respect to these combined
indices. We then expand $\rho^{\left(2\right)}$ with respect to its
eigenfunctions $\phi_{p,\alpha\beta}\left(\boldsymbol{r}_{1},\boldsymbol{r}_{2}\right)$:
\begin{equation}
\rho_{\alpha\beta\gamma\delta}^{\left(2\right)}\left(\boldsymbol{r}_{1},\boldsymbol{r}_{2};\boldsymbol{r}_{3},\boldsymbol{r}_{4}\right)=\sum_{p}n_{p}\phi_{p,\alpha\beta}^{*}\left(\boldsymbol{r}_{1},\boldsymbol{r}_{2}\right)\phi_{p,\gamma\delta}\left(\boldsymbol{r}_{3},\boldsymbol{r}_{4}'\right)
\end{equation}
with eigenvalues $n_{p}$. ODLRO is a state where the largest eigenvalue
$n_{0}$ is of the order of the particle number $N$. In this case
holds
\begin{equation}
\rho_{\alpha\beta\gamma\delta}^{\left(2\right)}\left(\boldsymbol{r}_{1},\boldsymbol{r}_{2};\boldsymbol{r}_{3},\boldsymbol{r}_{4}\right)\rightarrow n_{0}\phi_{0,\alpha\beta}^{*}\left(\boldsymbol{r}_{1},\boldsymbol{r}_{2}\right)\phi_{0,\gamma\delta}\left(\boldsymbol{r}_{3},\boldsymbol{r}_{4}\right)\label{eq:ODLR}
\end{equation}
in the limit where $\left|\boldsymbol{r}_{1,2}-\boldsymbol{r}_{3,4}\right|\rightarrow\infty$
while $\left|\boldsymbol{r}_{1}-\boldsymbol{r}_{2}\right|$ and $\left|\boldsymbol{r}_{3}-\boldsymbol{r}_{4}\right|$
remain finite. Hence, the long-distance physics of two-particle correlations
are dominated by the condensate \add{with condensate wave function $\phi_{0,\alpha\beta}\left(\boldsymbol{r}_{1},\boldsymbol{r}_{2}\right)$ (see fig.~\ref{fig:2pwf})}. If one analyses the BCS ground-state
wave function with gap $\Delta$ and density of states at the Fermi
level $\rho_{F}$, it follows that $n_{0}\sim\rho_{F}\left|\Delta\right|N$~\cite{Rensink1967},
as expected. ODLRO was also shown rigorously to occur in the negative-$U$
Hubbard model~\cite{Yang1989}, including in its ground state~\cite{Shen1993},
and in closely related models~\cite{Essler1992,Essler1993}. 

From the antisymmetry under the exchange of the operators $\psi_{\alpha}^{\dagger}\left(\boldsymbol{r}_{1}\right)\leftrightarrow\psi_{\beta}^{\dagger}\left(\boldsymbol{r}_{2}\right)$
and $\psi_{\gamma}\left(\boldsymbol{r}_{3}\right)\leftrightarrow\psi_{\delta}\left(\boldsymbol{r}_{4}\right)$
in $\rho^{\left(2\right)}$ follows that $\phi_{0,\alpha\beta}\left(\boldsymbol{r}_{1},\boldsymbol{r}_{2}\right)=-\phi_{0,\beta\alpha}\left(\boldsymbol{r}_{2},\boldsymbol{r}_{1}\right)$.
It has the properties of a two-particle fermion wave function. In
full analogy to the usual classification of anomalous expectation
values, see e.g. Ref.~\cite{Sigrist1991}, one can now expand 
\begin{equation}
\phi_{0,\alpha\beta}\left(\boldsymbol{r}_{1},\boldsymbol{r}_{2}\right)=\varphi_{0,s}\left(\boldsymbol{r}_{1},\boldsymbol{r}_{2}\right)i\sigma_{\alpha\beta}^{y}+\boldsymbol{\varphi}_{0,t}\left(\boldsymbol{r}_{1},\boldsymbol{r}_{2}\right)\cdot i\left(\boldsymbol{\sigma}\sigma^{y}\right)_{\alpha\beta}
\end{equation}
in terms of singlet and triplet contributions in spin space. Here
$\sigma^{l}$ stand for the Pauli matrices in spin space. Hence, all
our conclusions apply equally to singlet or triplet superconductors
or to combinations thereof as they occur in inversion symmetry breaking
systems. 

Important insights about the magnetic field behavior of superconductors,
such as the Meissner effect and flux quantization follow from ODLRO
because of a gauge argument. To see this we first consider a spatially homogeneous
magnetic field $\boldsymbol{B}={\rm const.}$~\cite{Sewell90,Nieh95} \add{and couple it to the charged fermions via minimal substitution. In particular this implies that the many-body wave function has to transform covariantly under local gauge transformations.}
The vector potential can be written as 
\begin{equation}
\boldsymbol{A}\left(\boldsymbol{r}\right)=\boldsymbol{A}_{0}\left(\boldsymbol{r}\right)+\nabla\varphi\left(\boldsymbol{r}\right),
\end{equation}
where $\boldsymbol{A}_{0}\left(\boldsymbol{r}\right)=\frac{1}{2}\boldsymbol{B}\times\boldsymbol{r}$
and $\varphi\left(\boldsymbol{r}\right)$ is an arbitrary function.
A spatial translation $\boldsymbol{r}\rightarrow\boldsymbol{r}-\boldsymbol{a}$
can be understood as a gauge transformation since the vector potential
transforms as 
\begin{eqnarray}
\boldsymbol{A}\left(\boldsymbol{r}\right) & \rightarrow & \boldsymbol{A}\left(\boldsymbol{r}+\boldsymbol{a}\right)=\boldsymbol{A}\left(\boldsymbol{r}\right)+\nabla\chi_{\boldsymbol{a}}\left(\boldsymbol{r}\right),\label{eq:gauge homogeneous}
\end{eqnarray}
with 
\begin{eqnarray}
\chi_{\boldsymbol{a}}\left(\boldsymbol{r}\right) & = & \boldsymbol{a}\cdot{\bf A}_{0}\left(\boldsymbol{r}\right)+\varphi\left(\boldsymbol{r}-\boldsymbol{a}\right)-\varphi\left(\boldsymbol{r}\right).\label{eq:gauge_phase}
\end{eqnarray}
\add{Since fermionic operators have to transform covariantly under local gauge transformations} it follows $\psi_{\alpha}\left(\boldsymbol{r}\right)=e^{i\frac{e}{\hbar c}\chi_{\boldsymbol{a}}\left(\boldsymbol{r}\right)}\psi_{\alpha}\left(\boldsymbol{r}-\boldsymbol{a}\right)$.
If the \add{system is translation symmetric, it follows from expressing the two-particle density matrix as an expectation value of fermion operators (Eq.~\eqref{eq:rho2}) that}
\begin{eqnarray}
\rho_{\alpha\beta\gamma\delta}^{\left(2\right)}\left(\boldsymbol{r}_{1},\boldsymbol{r}_{2};\boldsymbol{r}_{3},\boldsymbol{r}_{4}\right) & = & e^{-i\frac{e}{\hbar c}\left(\chi_{\boldsymbol{a}}\left(\boldsymbol{r}_{1}\right)+\chi_{\boldsymbol{a}}\left(\boldsymbol{r}_{2}\right)-\chi_{\boldsymbol{a}}\left(\boldsymbol{r}_{3}\right)-\chi_{\boldsymbol{a}}\left(\boldsymbol{r}_{4}\right)\right)}\nonumber \\
 & \times & \rho_{\alpha\beta\gamma\delta}^{\left(2\right)}\left(\boldsymbol{r}_{1}-\boldsymbol{a},\boldsymbol{r}_{2}-\boldsymbol{a};\boldsymbol{r}_{3}-\boldsymbol{a},\boldsymbol{r}_{4}-\boldsymbol{a}\right).\label{eq:2pgauge}
\end{eqnarray}
Without ODLRO this behavior of $\rho^{\left(2\right)}$ under gauge
transformations or translations does not allow to make strong statements
about the eigenfunctions $\phi_{p,\gamma\delta}\left(\boldsymbol{r}_{3},\boldsymbol{r}_{4}\right)$. We perform now two consecutive displacements by two non-collinear
vectors $\boldsymbol{a}_{1}$ and $\boldsymbol{a}_{2}$ in alternate
order. \add{For a general two-particle density matrix this leads to the condition
\begin{align}
    i\frac{e}{\hbar c} ( &\chi_{\bm{a}_1}(\bm{r}_1) - \chi_{\bm{a}_1}(\bm{r}_3) + \chi_{\bm{a}_2}(\bm{r}_1-\bm{a}_1) - \chi_{\bm{a}_2}(\bm{r}_3-\bm{a}_1) - \chi_{\bm{a}_2}(\bm{r}_1) +\chi_{\bm{a}_2}(\bm{r}_3) - \chi_{\bm{a}_1}(\bm{r}_1-\bm{a}_2) + \chi_{\bm{a}_1}(\bm{r}_3-\bm{a}_2) \nonumber\\ &\chi_{\bm{a}_1}(\bm{r}_2) - \chi_{\bm{a}_1}(\bm{r}_4) + \chi_{\bm{a}_2}(\bm{r}_2-\bm{a}_1) - \chi_{\bm{a}_2}(\bm{r}_4-\bm{a}_1) - \chi_{\bm{a}_2}(\bm{r}_2) +\chi_{\bm{a}_2}(\bm{r}_4) - \chi_{\bm{a}_1}(\bm{r}_2-\bm{a}_2) + \chi_{\bm{a}_1}(\bm{r}_4-\bm{a}_2)  ) \in \mathbb{Z}.
\end{align}
This condition is automatically fulfilled, since the left-hand side is identically zero.}
However, once we have a macroscopic condensate and can use Eq.~\ref{eq:ODLR}
it follows 
\begin{equation}
\phi_{0,\alpha\beta}\left(\boldsymbol{r}_{1},\boldsymbol{r}_{2}\right)=f_{\boldsymbol{a}}e^{i\frac{e}{\hbar c}\left(\chi_{\boldsymbol{a}}\left(\boldsymbol{r}_{1}\right)+\chi_{\boldsymbol{a}}\left(\boldsymbol{r}_{2}\right)\right)}\phi_{0,\alpha\beta}\left(\boldsymbol{r}_{1}-\boldsymbol{a},\boldsymbol{r}_{2}-\boldsymbol{a}\right),\label{eq:WF-gauge}
\end{equation}
where $f_{\boldsymbol{a}}$ is an $\boldsymbol{r}$-independent but
displacement dependent phase factor $\left|f_{\boldsymbol{a}}\right|=1$.
\add{And requiring successive displacements to commute} yields a condition on the phases
\begin{equation}
\chi_{\boldsymbol{a}_{2}}\left(\boldsymbol{r}\right)+\chi_{\boldsymbol{a}_{1}}\left(\boldsymbol{r}-\boldsymbol{a}_{2}\right)-\chi_{\boldsymbol{a}_{1}}\left(\boldsymbol{r}\right)-\chi_{\boldsymbol{a}_{2}}\left(\boldsymbol{r}-\boldsymbol{a}_{1}\right)=\frac{hc}{e}n \label{eq:quantization}
\end{equation}
with integer $n$. \add{Abstractly speaking, this condition is equivalent to the requirement that the projective representation of the group of translations Eq.~\eqref{eq:WF-gauge} preserves the commutativity of translations.} The expression Eq.~\ref{eq:gauge_phase} allows
to write condition \add{Eq.~\eqref{eq:quantization}} as:
\begin{equation}
\boldsymbol{B}\cdot\left(\boldsymbol{a}_{1}\times\boldsymbol{a}_{2}\right)=n\Phi_{0}.
\end{equation}
In the continuum, where any displacement $\boldsymbol{a}_{i}$ is
allowed, we can continuously change the left hand side of this equation.
Since the right hand side cannot be changed continuously, the only
solution is $\boldsymbol{B}=0$, which yields the Meissner effect
for homogeneous fields. In a periodic solid, the $\boldsymbol{a}_{i}$
must be integer multiples of the primitive lattice vectors. The smallest
non-zero field allowed would then have to place a flux quantum in
the unit cell of the system as discussed in Refs.~\cite{Tesanovic1989,Norman1991}. While this
excludes currently achievable fields for ordinary solids, this regime becomes relevant for moir\'{e}
materials, where the unit cells can be much larger. For a recent discussion of the
related Hofstadter superconductors, see Ref.~\cite{Shaffer2021}. 

Using Eq.~\ref{eq:WF-gauge} and considering a continuum description,
we can also perform an infinite sequence of infinitesimal displacements
along a path 
\begin{equation}
\phi_{0,\alpha\beta}\left(\boldsymbol{r}_{1}',\boldsymbol{r}_{2}'\right)=f_{\int_{\boldsymbol{r}_{1}}^{\boldsymbol{r}_{2}}d\boldsymbol{r}}e^{-i\frac{e}{\hbar c}\left(\int_{\boldsymbol{r}_{1}}^{\boldsymbol{r}_{1}'}\boldsymbol{A}\left(\boldsymbol{r}\right)\cdot d\boldsymbol{r}+\int_{\boldsymbol{r}_{2}}^{\boldsymbol{r}_{2}'}\boldsymbol{A}\left(\boldsymbol{r}\right)\cdot d\boldsymbol{r}\right)}\phi_{0,\alpha\beta}\left(\boldsymbol{r}_{1},\boldsymbol{r}_{2}\right).
\end{equation}
Here, the path that connects $\boldsymbol{r}_{1}$ with\textbf{ $\boldsymbol{r}_{1}'$}
must be the same as the one that connects $\boldsymbol{r}_{2}$ with\textbf{
$\boldsymbol{r}_{2}'$} . In the case of a closed loop follows 
\begin{equation}
\phi_{0,\alpha\beta}\left(\boldsymbol{r}_{1},\boldsymbol{r}_{2}\right)=e^{-i\frac{2e}{\hbar c}\oint\boldsymbol{A}\left(\boldsymbol{r}\right)\cdot d\boldsymbol{r}}\phi_{0,\alpha\beta}\left(\boldsymbol{r}_{1},\boldsymbol{r}_{2}\right).
\end{equation}
The result for the quantization of the flux follows from the single-valuedness of the wave function and  is
\begin{equation}
\Phi=\oint\boldsymbol{A}\left(\boldsymbol{r}\right)\cdot d\boldsymbol{r}=n\Phi_{0}.
\end{equation}
This analysis of Refs.~\cite{Yang62,Sewell90,Nieh95} reveals very
transparently that macroscopic coherence in fermionic systems, reflected
in a single large eigenvalue $n_{0}$ of $\rho^{\left(2\right)}$
of the order of the system size $N$, is the crucial ingredient that leads
to the Meissner effect and to flux quantization.  

\section{ODLRO and integer and fractional vortex lattice states}

\subsection{Integer flux vortex lattices}

In this section we allow for periodic magnetic fields subject to the
following properties: $\boldsymbol{B}$ points in the $z-$direction
and is periodic in the $xy$-plane, i.e. 
\begin{equation}
\boldsymbol{B}\left(\boldsymbol{r}\right)=\boldsymbol{B}\left(\boldsymbol{r}+\boldsymbol{a}_{i}\right),
\end{equation}
with $i=1,2$, where $\boldsymbol{a}_{1}$ and $\boldsymbol{a}_{2}$
that are both orthogonal to $\boldsymbol{e}_{z}$, the unit vector
along the $z$-direction; see Fig.~\ref{fig:Summary}. Moreover $\boldsymbol{B}\left(\boldsymbol{r}\right)$
shall be independent of the $z$-coordinate. The question is now,
what restriction does the presence of ODLRO pose on the magnetic field
configuration? We have already seen that for a homogeneous field that is not too large, the
only possible choice is a vanishing field and are now seeking to find
the corresponding restriction for a periodic field.

In the homogeneous case, a spatial translation of the system was recognized
with Eq.~\ref{eq:gauge homogeneous} as a gauge transformation. This
is physically transparent, since the magnetic field configuration
viewed from the displaced position is identical and therefore the
vector potential $\boldsymbol{A}$ can differ at most by a gauge transformation.
For a periodic field, this is only the case for a subset of translations,
namely the discrete lattice translations. To show this, let us calculate
explicitly the gauge transformation associated with such a displacement.
As the magnetic field is periodic it can be expanded in a Fourier
series using the reciprocal lattice vectors $\boldsymbol{K}$ of the
periodic field configuration. Then we can perform the Fourier expansion
\begin{equation}
\boldsymbol{B}\left(\boldsymbol{r}\right)=\sum_{\boldsymbol{K}}e^{i\boldsymbol{K}\cdot\boldsymbol{r}}B_{\boldsymbol{K}}\boldsymbol{e}_{z}.\label{eq:Vortex lattice}
\end{equation}
 We assume that $\boldsymbol{a}_{1,2}$ are multiples of the underlying
crystalline lattice. One can now explicitly generate a general expression
for the vector potential 
\begin{equation}
\boldsymbol{A}\left(\boldsymbol{r}\right)=\frac{1}{2}B_{\boldsymbol{0}}\boldsymbol{e}_{z}\times\boldsymbol{r}+\sum_{\boldsymbol{K}\neq0}e^{i\boldsymbol{K}\cdot\boldsymbol{r}}\frac{iB_{\boldsymbol{K}}}{\left|\boldsymbol{K}\right|}\boldsymbol{K}'+\nabla\varphi\left(\boldsymbol{r}\right),
\end{equation}
where $\boldsymbol{K}'$ is defined for every reciprocal lattice vector
$\boldsymbol{K}$ as the unique unit vector that satisfies $\frac{\boldsymbol{K}}{\left|\boldsymbol{K}\right|}\times\boldsymbol{K}'=-\boldsymbol{e}_{z}$.
The function $\varphi$ ensures that the gauge choice is still arbitrary.
Let us now show that a translation by a lattice vector $\boldsymbol{a}_{i}$
can be represented by a gauge transformation: 
\begin{eqnarray}
\boldsymbol{A}\left(\boldsymbol{r}-\boldsymbol{a}_{i}\right) & = & \frac{1}{2}B_{\boldsymbol{0}}\boldsymbol{e}_{z}\times\left(\boldsymbol{r}-\boldsymbol{a}_{i}\right)+\sum_{\boldsymbol{K}\neq0}e^{i\boldsymbol{K}\cdot\left(\boldsymbol{r}-\boldsymbol{a}_{i}\right)}\frac{iB_{\boldsymbol{K}}}{\left|\boldsymbol{K}\right|}\boldsymbol{K}'+\nabla\varphi\left(\boldsymbol{r}-\boldsymbol{a}_{i}\right)\nonumber \\
 & = & \boldsymbol{A}\left(\boldsymbol{r}\right)-\frac{1}{2}B_{\boldsymbol{0}}\boldsymbol{e}_{z}\times\boldsymbol{a}_{i}+\nabla\varphi\left(\boldsymbol{r}-\boldsymbol{a}_{i}\right)-\nabla\varphi\left(\boldsymbol{r}\right)\nonumber \\
 & = & \boldsymbol{A}\left(\boldsymbol{r}\right)+\nabla\chi_{\boldsymbol{a}_{i}}\left(\boldsymbol{r}\right),
\end{eqnarray}
with
\begin{equation}
\chi_{\boldsymbol{a}_{i}}\left(\boldsymbol{r}\right)=\frac{1}{2}B_{\boldsymbol{0}}\boldsymbol{a}_{i}\cdot\left(\boldsymbol{e}_{z}\times\boldsymbol{r}\right)+\varphi\left(\boldsymbol{r}-\boldsymbol{a}_{i}\right)-\varphi\left(\boldsymbol{r}\right).\label{eq:gauge_VL}
\end{equation}
We have used that for a lattice vector $\boldsymbol{a}_{i}$ holds
that $\boldsymbol{a}_{i}\cdot\boldsymbol{K}=2\pi k_i$ with $k_i\in\mathbb{Z}$.
As was shown in the previous section, the presence of ODLRO gives
rise to Eq.~\ref{eq:WF-gauge}. Consider again subsequent lattice translations
around the unit cell spanned by $\boldsymbol{a}_1$ and $\boldsymbol{a}_2$.
The condition that the wave function be single valued leads to 
\begin{equation}
f_{\boldsymbol{a}_1}f_{\boldsymbol{a}_2}e^{i\frac{2e}{\hbar c}\left(\chi_{\boldsymbol{a}_2}\left(\boldsymbol{r}-\boldsymbol{a}_1\right)+
\chi_{\boldsymbol{a}_1}\left(\boldsymbol{r}\right)\right)}=f_{\boldsymbol{a}_1}
f_{\boldsymbol{a}_2}e^{i\frac{2e}{\hbar c}\left(\chi_{\boldsymbol{a}_1}\left(\boldsymbol{r}-\boldsymbol{a}_2\right)+
\chi_{\boldsymbol{a}_2}\left(\boldsymbol{r}\right)\right)}.
\end{equation}
Since the $f_{\boldsymbol{a}_1}$ and $f_{\boldsymbol{a}_2}$ are just complex numbers
they can be cancelled and we obtain
\begin{equation}
e^{-i\frac{2e}{\hbar c}\left(\chi_{\boldsymbol{a}_1}\left(\boldsymbol{r}-\boldsymbol{a}_2\right)+\chi_{\boldsymbol{a}_2}\left(\boldsymbol{r}\right)-\chi_{\boldsymbol{a}_2}\left(\boldsymbol{r}-\boldsymbol{a}_1\right)-\chi_{\boldsymbol{a}_1}\left(\boldsymbol{r}\right)\right)}=1.
\end{equation}
Using $\chi_{\boldsymbol{a}_i}\left(\boldsymbol{r}\right)$ of Eq.~\ref{eq:gauge_VL}
we obtain for the combined gauge functions in the exponent
\begin{equation}
\chi_{\boldsymbol{a}_1}\left(\boldsymbol{r}-\boldsymbol{a}_2\right)+\chi_{\boldsymbol{a}_2}\left(\boldsymbol{r}\right)-\chi_{\boldsymbol{a}_2}\left(\boldsymbol{r}-\boldsymbol{a}_1\right)-\chi_{\boldsymbol{a}_1}\left(\boldsymbol{r}\right)=-B_{\boldsymbol{0}}\boldsymbol{e}_{z}\cdot\left(\boldsymbol{a}_1\times\boldsymbol{a}_2\right).
\end{equation}
This last expression has a clear physical meaning as minus the magnetic
flux that passes through the unit cell. To show this we use $\boldsymbol{B}\left(\boldsymbol{r}\right)$
of Eq.~\ref{eq:Vortex lattice} and determine 
\begin{eqnarray}
\Phi & = & \int\boldsymbol{B}\left(\boldsymbol{r}\right)\cdot d\boldsymbol{S}\nonumber \\
 & = & B_{\boldsymbol{0}}\boldsymbol{e}_{z}\cdot\left(\boldsymbol{a}_1\times\boldsymbol{a}_2\right)+\sum_{\boldsymbol{K}\neq0}B_{\boldsymbol{K}}\int e^{i\boldsymbol{K}\cdot\boldsymbol{r}}\boldsymbol{e}_{z}\cdot d\boldsymbol{S}.
\end{eqnarray}
One easily sees that the second term vanishes since the integration
is over the parallelogram spanned by lattice vectors $\boldsymbol{a}_{1}$
and $\boldsymbol{a}_{2}$ with $d\boldsymbol{S}\propto\boldsymbol{a}_{1}\times\boldsymbol{a}_{2}$.
Therefore we obtain

\begin{equation}
e^{i\frac{2e}{\hbar c}\Phi}=1,
\end{equation}
which implies that the flux through a unit cell of the lattice is
quantized in integer units of the flux quantum $\Phi_{0}$. In other
words, a vortex lattice with integer flux in each unit cell is the only static, periodic field configuration of the type discussed
above that is consistent with off-diagonal long-range order. 

\subsection{Fractionalized vortex lattices}

So far an implicit assumption for ODLRO has been that the largest
eigenvalue $n_{0}$ of the two-particle density matrix in Eq.~\ref{eq:ODLR}
is unique. Next we address what happens when there are $g$ degenerate
eigenstates $\phi_{0,\alpha\beta}^{\left(i\right)}\left(\boldsymbol{r}_{1},\boldsymbol{r}_{2}\right)$ with $i=1,\cdots, g$
of the two-particle density matrix $\rho^{\left(2\right)}$ . For the
long distance behavior $\left|\boldsymbol{r}_{1,2}-\boldsymbol{r}_{3,4}\right|\rightarrow\infty$
with $\left|\boldsymbol{r}_{1}-\boldsymbol{r}_{2}\right|$ and $\left|\boldsymbol{r}_{3}-\boldsymbol{r}_{4}\right|$
finite, it follows now
\begin{equation}
\rho_{\alpha\beta\gamma\delta}^{\left(2\right)}\left(\boldsymbol{r}_{1},\boldsymbol{r}_{2};\boldsymbol{r}_{3},\boldsymbol{r}_{4}\right)\rightarrow n_{0}\sum_{i=1}^{g}\phi_{0,\alpha\beta}^{\left(i\right)*}\left(\boldsymbol{r}_{1},\boldsymbol{r}_{2}\right)\phi_{0,\gamma\delta}^{\left(i\right)}\left(\boldsymbol{r}_{3},\boldsymbol{r}_{4}\right).\label{eq:ODLR-1}
\end{equation}
We arrange these eigenstates in the $g$-component vector $\boldsymbol{\phi}_{0,\alpha\beta}\left(\boldsymbol{r}_{1},\boldsymbol{r}_{2}\right)$
such that
\begin{equation}
\rho_{\alpha\beta\gamma\delta}^{\left(2\right)}\left(\boldsymbol{r}_{1},\boldsymbol{r}_{2};\boldsymbol{r}_{3},\boldsymbol{r}_{4}\right)\rightarrow n_{0}\boldsymbol{\phi}_{0,\alpha\beta}^{*}\left(\boldsymbol{r}_{1},\boldsymbol{r}_{2}\right)\cdot\boldsymbol{\phi}_{0,\gamma\delta}\left(\boldsymbol{r}_{3},\boldsymbol{r}_{4}\right).
\end{equation}
The generic behavior Eq.~\ref{eq:2pgauge} of $\rho^{\left(2\right)}$
under gauge transformations is of course unchanged. With multi-component
ODLRO we then obtain the following transformation behavior of the
wave function under translations by a lattice vector $\boldsymbol{a}_{i}$
\begin{equation}
\boldsymbol{\phi}_{0,\alpha\beta}\left(\boldsymbol{r}_{1},\boldsymbol{r}_{2}\right)=e^{i\frac{e}{\hbar c}\left(\chi_{\boldsymbol{a}_{i}}\left(\boldsymbol{r}_{1}\right)+\chi_{\boldsymbol{a}_{i}}\left(\boldsymbol{r}_{2}\right)\right)}\hat{f}_{\boldsymbol{a}_{i}}\cdot\boldsymbol{\phi}_{0,\alpha\beta}\left(\boldsymbol{r}_{1}-\boldsymbol{a}_{i},\boldsymbol{r}_{2}-\boldsymbol{a}_{i}\right),
\end{equation}
where we consider again a magnetic field periodic in the $xy$-plane
and independent on the $z$-coordinate. $\hat{f}_{\boldsymbol{a}}$,
which was formerly a phase factor, is now a unitary $g\times g$ matrix.
It expresses the fact that the choice of basis at each point in space
is arbitrary. Different components of the order parameter mix under
gauge transformations and translations. If we now use the gauge function
of Eq.~\ref{eq:gauge_VL} for periodic field configurations, the single-valuedness
of the eigenfunctions implies
\begin{equation}
\hat{f}_{\boldsymbol{a}_{1}}\cdot\hat{f}_{\boldsymbol{a}_{2}}=\hat{f}_{\boldsymbol{a}_{2}}\cdot\hat{f}_{\boldsymbol{a}_{1}}e^{i\frac{2e}{\hbar c}\Phi}.
\end{equation}
where $\Phi$ is again the flux through the parallelogram spanned by
$\boldsymbol{a}_{1,2}$. Taking the determinant of this expression
on both sides and using $\det\hat{f}_{\boldsymbol{a}_{i}}\cdot\hat{f}_{\boldsymbol{a}_{j}}=\det\hat{f}_{\boldsymbol{a}_{i}}\det\hat{f}_{\boldsymbol{a}_{j}}$
and $\det\left(e^{i\alpha}\hat{f}_{\boldsymbol{a}_{i}}\right)=e^{ig\alpha}\det\hat{f}_{\boldsymbol{a}_{i}}$,
finally yields $e^{ig\frac{2e}{\hbar c}\Phi}=1$ . This leads to the
quantization condition
\begin{equation}
\Phi=\frac{n}{g}\Phi_{0}.
\label{fracVL}
\end{equation}
The magnetic flux through the unit cell of the vortex lattice is thus
quantized in fractional values of the flux quantum, where the denominator
is given by the degree of degeneracy $g$. 

One has to be somewhat careful with this argument. Strictly speaking, we find that such fractionalized vortices cannot be excluded if one only considers the determinant of the above condition Eq.~\ref{fracVL} and there could be other, more stringent conditions in the full equation. An example, where we can confirm Eq.~\ref{fracVL} is a $g$-component  order parameter manifold that transforms like $U(g)=U(1)\times SU(g)$. The $g$ eigenfunctions of the two-particle density matrix introduced above in fact transform under an irreducible representation of this group. The presence of stable line defects with a quantized integer index is guaranteed by the fundamental group being isomorphic to the integers $\pi_1(U(g))\simeq\mathbb{Z}$~\cite{Mermin1979}. In a $1$-component condensate these are vortices carrying an integer multiple of the flux quantum. Generally, the fundamental group does not carry any information about the physical meaning of the quantized index. The connection of the index to observable quantities has to be provided by identifying the properties of the actual defects. In the case of flux quantization we know that only the global phase couples to the electromagnetic field and that the trapped flux is proportional to the winding number. Thus, by identifying the fundamental defect and computing the winding number of the global phase the unit of flux quantization can be found.

Take as an example a two-fold degenerate state, i.e. $\hat{f}_{\boldsymbol{a}}\in U(2)$. A general element $\hat{f}$ of $U(2)$ can be written as
\begin{equation}
\hat{f} = e^{i\theta}(n_0 1 + \boldsymbol{n}\cdot\boldsymbol{\sigma}), \quad n_0^2+\boldsymbol{n}^2=1.
\end{equation}
Parametrizing the defect by $\phi\in[0,2\pi)$ we can construct
\begin{equation}
\hat{f} = e^{i\phi/2}(\cos\frac{\phi}{2} 1 + \sin\frac{\phi}{2}\sigma^z).
\end{equation}
This corresponds to a defect that carries one half of a flux quantum, because the global phase winds by $1/2$. Essentially, this is possible because $(-1)\in SU(2)$. We can generalize this by noting that $e^{i\frac{2\pi}{g}(g-1)}1\in SU(g)$. The fundamental defect can be constructed by connecting $1$ and $e^{i\frac{2\pi}{g}(g-1)}1$ in $SU(g)$ and simultaneously connecting $1$ and $e^{i\frac{2\pi}{g}}$ in $U(1)$. The former  is always possible, because $SU(g)$ is simply connected. The resulting loop cannot be deformed to a point in $U(g)$ and it carries a flux of $\Phi_0/g$. This is fully consistent with  Eq.~\ref{fracVL}. Notice, this conclusion relies on our assumption that the order parameter  of the problem transforms under $U(g)$. Other order-parameter manifolds require their own, but analogous analysis.

\section{Summary}

In summary, we generalized the implications of off-diagonal long-range
order in superconductors to periodically inhomogeneous magnetic fields.
For single component superconductors one finds that a condition for
the existence of finite fields is that the flux per unit cell is a
multiple of the elementary flux quantum. This is of course the established Abrikosov vortex lattice.
Still, our derivation has the appeal that it is valid   for situations
where the BCS or Ginzburg-Landau theories of superconductivity may not apply. Moreover,
the rather simple nature of the proof may be of some appeal on its
own right. The generalization to multi-component superconductors is relevant whenever the order parameter transforms according
to a higher-dimensional irreducible representation of the symmetry group.  
It shows that now fractionalized vortex lattices become generally allowed  inhomogeneous magnetic field states.  
\begin{acknowledgments}
We are grateful to R. Willa for helpful discussions and acknowledge support by the Deutsche Forschungsgemeinschaft (German Research Foundation) Project ER 463/14-1. We acknowledge support by the KIT-Publication Fund of the Karlsruhe Institute of Technology.
\end{acknowledgments}

\end{document}